\input harvmac

\Title{HUTP-98/A060}
{\vbox{\hbox{Neutrinos on Earth}\vskip5pt
\hbox{and in the Heavens}}}
\centerline{\bf Howard Georgi\footnote{$^*$}{Research supported in part by the
National Science Foundation
under grant number NSF-PHY/98-02709.}%
\footnote{$^1$}{georgi@physics.harvard.edu}
\&\ S.L. Glashow$^*$\footnote{$^2$}{glashow@physics.harvard.edu} }\medskip
\vbox{\it \centerline{Laboratory of Physics}\vskip-2pt
\centerline{Harvard University}\vskip-2pt
\centerline{Cambridge, MA 02138}}

\bigskip
\bigskip
\centerline{Abstract}

\medskip
{\baselineskip=10pt\centerline{\vbox{\hsize=.8\hsize\ninerm\noindent 
Recent data lead us to a simple and intriguing form of the neutrino
mass matrix. In particular, we find  solar neutrino oscillations to
be nearly maximal (and rule out the small-angle MSW explanation of
solar neutrino observations) if relic neutrinos comprise at least 
one~percent of the critical mass density of the universe. 
}}}

\Date{08/98}
%\draft
\baselineskip=12pt

Cosmologists differ on whether or not neutrinos play an essential role in
the evolution of the large-scale structure of the universe~\ref\rpri{For a
recent review, see J.R. Primack, ``Dark Matter and Structure Formation,''
SCIPP-96-59-REV, Jul 1997, to be
published in the proceedings of Midrasha Mathematicae in Jerusalem: Winter
School in
Dynamical Systems, Jerusalem, Israel, 12-17 Jan 1997, 
{\tt astro-ph/9707285.}}. In this paper, we assume that they do,
and
that the sum of their masses is several electron volts so that relic
neutrinos comprise several percent of the critical mass density.  Under
this hypothesis, we demonstrate how a wide range of observations and
deductions relating to neutrinos can be explained in terms of a specific
effective neutrino mass matrix $\cal M$ involving a suggestive pattern of
neutrino masses and mixings. Although many of our arguments may be found
elsewhere in part or in other contexts~\ref\rothers{A good recent review
is R. N. Mohapatra, {\tt hep-ph/9808284}. Other recent papers on the
subject include: J.W.F. Valle, {\tt
hep-ph/9509306}; H. Minikata and O. Yasuda, Nucl. Phys. {\bf B523} (1997)
597; F. Vissani, {\tt hep-ph/9708473}; H. Minakata and O. Yasuda, Phys.
Rev. {\bf D56} (1997) 1692.}, a cogent synthesis may be useful.

The literature is rife with both experimental and theoretical claims
regarding neutrino properties, many of them in conflict with one another.
Below is the somewhat arbitrary selection of
neutrino `facts' we shall accept and describe. 
These facts are consistent with one another and 
are suggested by current experimental data, but they are not decisively
established. We show how this particular set of facts
constrains the neutrino mass matrix to have a form that we find both
fascinating and a bit
bizarre~\ref\rbarger{It is in fact the `bimaximal mixing matrix' described by
V. Barger, S. Pakvasa, T.J. Weiler, and K.
Whisnant, MADPH-98-1063, June 1998, {\tt hep-ph/9806387}.}.

\medskip

\item{F1.} There exist precisely three chiral neutrino states with Majorana
masses, $m_1$, $m_2$ and $m_3$ (taken to be real and non-negative). In
particular we do not consider the existence of additional neutrinos,
sterile or otherwise.

\item{F2.} Atmospheric neutrinos rarely oscillate into electron neutrinos. 
This
is a plausible, but not inescapable~\ref\rbarbieri{See: R. Barbieri, {\it et
al.},
Oscillations of Solar and Atmospheric Neutrinos, IFUP-TH-25-98, Jun 1998, 
{\tt hep-ph/9807235.} },
interpretation of recent data from the Super-Kamiokande
Collaboration~\ref\rsk{T. Kajita, to appear in Proc. XVIII
Intern'l Conf. on Neutrino Physics and Astrophysics, Takayama,
Japan, June 1998; Super-Kamiokande Collaboration, {\tt hep-ex/9807003.}}
and  from CHOOZ~\ref\rchooz{M. Apollonio {\it et al.,} Phys. Lett. {\bf
B420} (1998) 397.}.

\item{F3.} Atmospheric muon neutrinos suffer maximal, or
nearly maximal, two-flavor
oscillations into tau neutrinos~\rsk. The relevant mixing angle
satisfies:
\eqn\etheta{\sin^2{2\theta}>0.82\,,}
and the required neutrino mass-squared difference $\Delta_a$
 satisfies:
\eqn\edeltaa{5\times 10^{-4}\,{\rm eV}^2<
\Delta_a < 6\times 10^{-3}\,{\rm eV}^2\,.}

\item{F4.} Oscillations are needed to resolve the discrepancy between the
observed and computed solar neutrino fluxes~\ref\rys{{E.g.,} Y. Suzuki,
to appear in Proc. XVIII Intern'l Conf.
on Neutrino Physics and Astrophysics,
Takayama, Japan, June 1998.}~\ref\rjb{{\it E.g.,} J.N. Bahcall,
{\tt hep-ph/9807216}.}. A relevant 
neutrino squared-mass difference  in the range:
\eqn\edeltas{6\times 10^{-11}\,{\rm eV}^2
 < \Delta_s < 2\times 10^{-5}\,{\rm eV}^2\,,}
provide MSW explanations for
larger values of $\Delta_s$, and just-so explanations for smaller values.
It has been suggested~\ref\rsnuss{S. Nussinov, Phys. Lett {\bf B63} (1976)
201; See \rjb\ for other references and for a critique
of this possibility.}\
that the solar neutrino deficit may result from maximal
time-averaged vacuum oscillations. If so,
the bound $\Delta_s< 10^{-3}\;
{\rm eV}^2$ is obtained from reactor experiments~\rchooz.
It is premature and unnecessary for us to choose amongst these
proposed solutions to the solar neutrino puzzle.

\item{F5.} Here we assume that neutrino masses are
large enough to play a significant cosmological role and take
\eqn\ecmass{
{m_1+m_2+m_3\over 3} \equiv M\sim 2~\rm eV\,.}
This is the least well established fact in our list, but it
is crucial to our discussion.

\item{F6.} Careful studies of many nuclear species have failed to detect
neutrinoless double beta decay. These experiments provide bounds on ${\cal
M}_{ee}$, the $ee$ component of the Majorana neutrino mass matrix in the
charged lepton flavor basis --- a weighted average of neutrino masses.
Currently,
the strongest bound is ${\cal M}_{ee}<B=0.46$~eV~\ref\rbb{L. Baudis {\it
et al.,}, Phys. Lett. {\bf B407} (1997) 219.}.

\bigskip

These `facts' dramatically constrain the form of the neutrino mass
matrix. In addition, they compel solar neutrino oscillations to be maximal,
thereby ruling out the small-angle MSW solution of the solar neutrino
puzzle.
We begin by considering the well-known implications of F1.
The most general mass matrix involving three chiral neutrinos
is a $3\times 3$
complex symmetric matrix $\cal M$. It may be written:
\eqn\em{{\cal M} = e^{i\eta}\, U_0^* D_0\, U_0^\dagger\,,}
where $U_0$ is an element of $SU(3)$ and
$D$ is a diagonal matrix with real non-negative entries $m_i$.
The mass matrix would be real were CP conserved, but it is
not. Consequently $\cal M$
involves nine convention-independent parameters.
Judicious choice of the phases of the flavor eigenstates allows us to
rewrite Eq.\em\ as:
\eqn\emm{ {\cal M} = U^* D\, U^\dagger}
where $U$ is a unitary `Kobayashi-Maskawa' matrix (involving three angles
$\theta_i$ and a complex phase $\delta$) expressing flavor eigenstates in
terms of mass eigenstates. In a standard notation:
\eqn\ekm{\pmatrix{\nu_e \cr \nu_\mu\cr \nu_\tau\cr}=
\pmatrix{c_2c_3& c_2s_3& s_2e^{-i\delta}\cr
-c_1s_3-s_1s_2c_3e^{i\delta}&
+c_1c_3-s_1s_2s_3e^{i\delta}&s_1c_2\cr
+s_1s_3-c_1s_2c_3e^{i\delta}&
-s_1c_3-c_1s_2s_3e^{i\delta}& c_1c_2\cr}\,
\pmatrix{\nu_1\cr \nu_2\cr \nu_3\cr}\,,}
with $s_i$ and $c_i$ standing for sines and cosines of $\theta_i$.
The remaining five parameters appear in the diagonal matrix $D$,
which may be written:
\eqn\ed{D=\pmatrix{m_1e^{i\phi}&0&0\cr 0&m_2e^{i\phi'}&0\cr 0&0&m_3\cr}\,.}
Each of the phase factors ($e^{i\delta},\, e^{i\phi},\,e^{i\phi'}$), if
not real, is CP violating.

The amplitude for atmospheric muon neutrinos
with energy $E_a$ to oscillate into $\nu_e$ over a distance $R_a$ is
\eqn\nutoe{\sum_j\,U_{\mu j}\,U_{ej}^*\,e^{im_j^2R_a/2E_a}\;.
}
According to fact F2, this amplitude
must be small over the range of $R_a$ and $E_a$
relevant to atmospheric neutrinos, around $2E_a/R_a\approx10^{-3}$~eV$^2$.
It follows that
$|m_j^2-m_k^2|\,R_a/2E_a$ must be small for some pair of neutrino mass
eigenstates $j$ and $k$. To prove this, we assume the contrary. It follows
that the
amplitude \nutoe\ is small for a range $R_a$ and $E_a$ if and only if
$U_{\mu j}\,U_{ej}^*$ is small for each $j$. But fact F3 requires
that $\nu_\mu$ is not close to a mass eigenstate. Thus $U_{\mu
j}\,U_{ej}^*$ can be small for each $j$ only if $\nu_e$ is close to a mass
eigenstate. This would lead us to the
the so-called small-angle
MSW solution that requires 
$\Delta_s < 2\times10^{-5}$~eV$^2$: small compared to $2E_a/R_a$
and contrary to the hypothesis --- QED.

Thus the neutrino mass
eigenstates associated with 
atmospheric oscillations must have a squared-mass difference 
$\Delta_a>5\times 10^{-4}\;\rm eV^2$, while those associated with
solar oscillations must have a much smaller
squared-mass difference, $\Delta_s\ll\Delta_a$. 
Without
loss of generality, we 
take $\Delta_a\equiv \vert m_3^2-m_2^2\vert$ and $\Delta_s\equiv \vert
m_2^2-m_1^2\vert$. 
This may
all sound rather obvious, and indeed it is the standard wisdom. But notice
that the argument of the previous paragraph rules out the  possibility of
the time-averaged  solar neutrino solution with $\Delta_s$ 
near the CHOOZ bound.

We showed  that all differences of squares of neutrino masses are less than
$10^{-3}\,\rm eV^2$. Yet according to F5 (and Eq.\ecmass\ in particular) the
sum of the neutrino masses must be several eV.\foot{The bound $M<4.4$~eV
follows from a recent measurement of the tritium beta
spectrum~\ref\rtrit{A.I. Belesev {\it et al.,} Phys. Lett. {\bf B350} (1995)
263.}.}
Thus the three neutrinos must
have equal mass to a precision of at least $10^{-3}$ if they are to play a
role in large-scale structure formation. From the point of view of particle
theory, this is truly a bizarre result and not at all what one would expect
from the simplest see-saw mechanisms. It is nonetheless an immediate
consequence of the facts we have accepted. We now proceed to a more detailed
discussion of the mass matrix \emm, from which additional constraints
can be found.

We may express the {\it in vacua\/} energy-dependent survival probabilities
for solar and atmospheric neutrinos in terms of the parameters so defined.
Because the path length $R_s$ of a solar neutrino is roughly an astronomical
unit, Eq.\edeltaa\ yields $\Delta_a R_s/E\gg 1$. Using this relation, we
obtain:
\eqn\esosc{
P(\nu_e\rightarrow\nu_e)\big\vert_{\rm solar}\simeq
1-{\sin^2{2\theta_2}\over 2}-
\cos^4{\theta_2}\sin^2{2\theta_3}\,\sin^2{(\Delta_s R_s/4E)}\,.}
Because the path length of an atmospheric neutrino
$R_a$ can be no greater than
Earth's diameter, Eq.\edeltas\ yields
$\Delta_s R_a/E\ll 1$. Using this relation we obtain:
\eqn\eaosc{
P(\nu_\mu\rightarrow\nu_\mu)\big\vert_{\rm atmospheric}=
1-4\sin^2{\theta_1}\cos^2{\theta_2}\,(1-
\sin^2{\theta_1}\cos^2{\theta_2})
\,\sin^2{(\Delta_a R_a/4E)}\,.}
These oscillations produce electron or tau neutrinos in the
ratio:
\eqn\eratio{{P(\nu_\mu\rightarrow\nu_e)\over P(\nu_\mu\rightarrow\nu_\tau)}
\bigg\vert_{\rm atmospheric}\simeq
{\tan^2{\theta_2}\over\cos^2{\theta_1}}\,.}
Note that none of the Eqs. \esosc, \eaosc, and \eratio\ involve the
CP-violating parameter $\delta$.

We turn to the consequences of our other tentatively accepted facts.
F2 and Eq.\eratio\ yield:
\eqn\eff{\theta_2\simeq 0\;,}
expressing the absence of oscillations of atmospheric oscillations into
electron neutrinos. This result greatly simplifies Eqs.\esosc\ and
\eaosc, which become:
\eqn\easosc{P\big\vert_{\rm solar}\simeq 1-\sin^2{2\theta_3}\sin^2{(\Delta_s
R_s/4E)},\quad
 P\big\vert_{\rm atmospheric}\simeq
1- \sin^2{2\theta_1}\sin^2{(\Delta_a R_a/4E)}\,.}
Thus $\theta_1$ is the parameter controlling atmospheric neutrino
oscillations, and we conclude from
F3 and Eq. \etheta\ that:
\eqn\efff{\sin{2\theta_1}\simeq 1\;,}
expressing the observation that these oscillations
are nearly maximal.

Finally, we must address F6, the observed suppression of
neutrinoless double beta decay. The amplitude for this process is
proportional to the quantity ${\cal M}_{ee}$, on which the bound is:
\eqn\ebb{{\cal M}_{ee}\equiv
\big\vert m_1\,c_2^2c_3^2e^{i\phi} + m_2\,c_2^2s_3^2e^{i\phi'} + m_3\,
s_2^2\,e^{i2\delta}\big\vert < B \,.}
We have seen that F5 requires
neutrino masses to be equal to a precision sufficient
to neglect their differences in Eq.\ebb. Furthermore, Eq.\eff\ lets
us put Eq.\ebb\ into the simple form:
\eqn\ebbb{\big\vert \cos^2{\theta_3}\,e^{i\phi} + \sin^2{\theta_3}\,e^{i\phi'}
\big\vert <B/M\;.}
Setting $M=2$~eV and $B=0.46$~eV (the current experimental upper
bound), we find that Eq.\ebb\ can be satisfied if $\phi+\phi'\simeq \pi$
and $\vert\cos{2\theta_3}\vert < 0.23$, or
\eqn\ethree{\sin^2{2\theta_3}> 0.95\,.}
Our six facts are mutually consistent if and only if
solar neutrino oscillations are nearly maximal. 
Somewhat stronger bounds on 
neutrinoless double beta decay could strengthen Eq.\ethree\ enough to leave
just-so oscillations~\ref\rjust{V. Barger, K. Whisnant, and R.J.N Phillips, 
Phys. Rev. {\bf D24} (1981) 538\semi S.L Glashow and L.M.  Krauss, Phys. Lett.
{\bf B190} (1987) 199.}~as the only viable explanation of the solar
neutrino data~ \rjb. 
Conversely, if the small
angle MSW description of solar neutrino oscillations is correct, the sum of
the neutrino masses is bounded above by $3B\simeq 1.4$~eV. In this case,
future double beta-decay experiments may exclude the
cosmological relevance of relic neutrinos.
\medskip

The neutrino mass matrix we are led to has approximately
the following form:
\eqn\mm{
{\cal M}=M\,\pmatrix{
0&{1\over\sqrt2}&{1\over\sqrt2}\cr
{1\over\sqrt2}&{1\over2}&-{1\over2}\cr
{1\over\sqrt2}&-{1\over2}&{1\over2}\cr
}}
This has several intriguing
properties: ${\cal MM}^\dagger$ is approximately a multiple of the
unit matrix and ${\cal M}_{ee}\simeq 0$. To the extent that these relations
are satisfied, the following processes are forbidden at all orders in
perturbation theory:

{\obeylines \medskip
Neutrinoless double beta decay
Muon decay into $e+\gamma$ or into $e+e+\bar e$
Muonium-antimuonium transitions
Muon--electron conversion via capture
The induction of an electron electric dipole moment
\medskip}
\noindent This point is academic because the
detection of any of
the above processes (except neutrinoless double beta decay)
would require a radical revision of the standard model.

Neutrino astronomy is a new science. Future observations of neutrinos from
nearby supernov\ae, or among ultra-high energy cosmic rays, are likely
sources of new information about particles and the universe. These
neutrinos, having traversed great distances, will experience time-averaged
oscillations so that their composition at detection will not coincide with
their composition at birth. Let $D_\ell$ be the number of detected
neutrinos with identities $\nu_\ell$, and
$B_\ell$
their numbers at birth.
With the above mixing parameters we find:
\eqn\enc{\pmatrix{D_e\cr D_\mu\cr D_\tau\cr }
={1\over 8}\,\pmatrix{4&2&2\cr 2&3&3\cr 2&3&3\cr}\,
\pmatrix{B_e\cr B_\mu\cr B_\tau\cr }}
Half of the $\nu_e$ burst from a supernova
reach Earth as $\nu_e$, while cosmic $\nu_\mu$'s are seen as
25\%\ $\nu_e$'s and 37.5\%\ $\nu_\tau$'s.
\medskip

Let us summarize our results. There are nine parameters in the neutrino
mass matrix, all but one of which are severely constrained by the facts we
have accepted. The three neutrino masses are nearly (but not quite!)
the same. The angles relating flavor and mass eigenstates take the
following simple values: $\theta_1\simeq\theta_3\simeq \pi/4$ and
$\theta_2\simeq 0$. 
These simple relations may indicate a deeper underlying
truth. 

In this connection, note that just one of the three {\it a priori\/}
CP-violating parameters in $\cal M$ is unconstrained by our analysis: We
must have $\phi-\phi'\simeq \pi$ to suppress neutrinoless double beta decay
and the parameter $\delta$ is {\it hors de combat\/} because it always
occurs multiplied by $\sin{\theta_2}$, which nearly vanishes.
 Perhaps the neutrino mass matrix is
real and CP conserving to lowest order in an unspecified underlying theory.
In that case and to that order, it is not implausible that the
Kobayashi-Maskawa matrix relevant to quarks is also real. 
We have come upon just such a result
in a model where CP breaking is both soft and
superweak~\ref\rsghg{H. Georgi and S.L Glashow, {\tt
hep-ph/9807399}.}.

\bigskip\bigskip
\centerline{\bf Acknowledgements}\medskip

One of us (SLG) thanks Maurice Goldhaber, Lawrence Krauss, and Lisa Randall
for stimulating discussions. We are also grateful to John Bahcall, Ed Kearns
and Stephen Parke for comments.
This work was supported in part by the National Science Foundation
under grant number NSF-PHY/98-02709.

\listrefs

\bye